\documentclass[%
aip,
amssymb,
reprint,
]{revtex4-1}
\usepackage{bm}
\usepackage{mathptmx}
\usepackage{etoolbox}
\usepackage{chemformula} 

\usepackage[T1]{fontenc} 
\usepackage{color}
\usepackage[utf8]{inputenc}
\usepackage{ifthen}
\usepackage{amsmath}
\usepackage{listings}
\usepackage{placeins}
\usepackage{float}
\usepackage{physics}
\usepackage{comment}
\usepackage{soul}
\usepackage[inline]{enumitem}
\usepackage{braket}
\usepackage{hyperref}
\usepackage{xcolor}
\usepackage{graphicx}
\usepackage{psfrag}
\usepackage{pgffor}
\usepackage{dcolumn}
\usepackage{array,ragged2e}
\usepackage{bbold}

\newcommand{\Bn}{\mathbf{n}}
\newcommand{\Be}{\mathbf{e}}

\newcommand{\Reals}{\mathbb{R}}

\makeatletter
\def\@email#1#2{%
 \endgroup
 \patchcmd{\titleblock@produce}
  {\frontmatter@RRAPformat}
  {\frontmatter@RRAPformat{\produce@RRAP{*#1\href{mailto:#2}{#2}}}\frontmatter@RRAPformat}
  {}{}
}%
\makeatother
\begin{document}

\preprint{AIP/123-QED}

\title{Extended Non-Markovian Stochastic Schr\"odinger Equation with Complex Frequency Modes for General Basis Functions}

\author{Yukai Guo}
\author{Zeyu Huang}
\author{Xing Gao}
\email{gxing@mail.sysu.edu.cn}
\affiliation{School of Materials, Sun Yat-sen University, Shenzhen, Guangdong 518107, China}

\date{\today}

\begin{abstract}
We introduce an extended formulation of the non-Markovian stochastic Schr\"odinger equation with complex frequency modes (extended cNMSSE), designed for simulating open quantum system dynamics under arbitrary spectral densities. 
This extension employs non-exponential basis sets to expand the bath correlation functions, overcoming the reliance of the original cNMSSE on exponential decompositions of the spectral density. Consequently, the extended cNMSSE is applicable to environments beyond those characterized by Debye-type spectral densities. 
The flexibility to employ general basis functions is particularly advantageous for handling spectral densities with higher-order poles, for which exponential decompositions are often inaccurate or unavailable.
The extended cNMSSE is implemented in a pseudo-Fock space using conventional ladder operators and solved efficiently via matrix product state (MPS) techniques, preserving the favorable linear-scaling and wavefunction-based nature of the original method.
Benchmark simulations across four representative cases, including discrete spectral density, Ohmic spectral density with exponential and algebraic cutoffs, and critically damped Brownian spectral density, demonstrate excellent agreement with results of hierarchy of forward-backward stochastic Schr\"odinger equations (HFB-SSE) and extended hierarchical equation of motion (HEOM).
\end{abstract}

\maketitle

\section{\label{introduction}Introduction}
Open quantum systems (OQS) are quantum mechanical systems that interact with external environments. In numerous physical and chemical processes, it is essential to account for non-Markovianity, where environmental memory effects play a significant role.\cite{breuer2002theory,weiss2012quantum,oliver2008charge} Addressing such dynamics has prompted the development of a variety of methods, including quasi-adiabatic propagator path integral (QUAPI),\cite{makri1995quapi_I,makri1995quapi_II} generalized quantum master equation (GQME),\cite{shi2003GQME,rabani2011GQME} hierarchical equations of motion (HEOM),\cite{tanimura1989time,ishizaki2005quantum} dissipative equation of motion (DEOM),\cite{yan2016deom_reivew} dissipaton-embedded quantum master equation in second quantization (DQME-SQ)\cite{li2024towards} and pseudomode theory.\cite{tamascelli2018nonperturbative,park2024quasi}
These methods are formulated within the density matrix framework. In contrast, the non-Markovian stochastic Schr\"odinger equation (NMSSE) provides an alternative wavefunction-based approach, offering favorable linear scaling with system size.

Non-Markovian quantum state diffusion (NMQSD) was originally proposed by Di\'osi and Strunz\cite{diosi1997non, diosi1998non} to describe the dynamics of OQS beyond the Markovian approximation. 
The primary technical difficulty in the NMQSD formalism lies in the treatment of the functional derivative, which has since been addressed through the development of the hierarchy of pure states (HOPS) using coherent state representations\cite{suess2014hierarchy} or path integral approaches.\cite{ke2016hierarchy,song2016alternative,ke2017perturbation}
Recently, we reformulated HOPS in a pseudo-Fock space with conventional ladder operators and wavefunction rescaling approaches, resulting in a non-Markovian stochastic Schr\"odinger equation with complex frequency modes (cNMSSE).\cite{gao2022non} The cNMSSE has been successfully solved within matrix product state (MPS) framework and applied to investigate charge transport in large model systems, incorporating both diagonal~\cite{gao2022non} and off-diagonal~\cite{zhou2024stochastic} electron-phonon coupling effects, as well as the non-adiabatic dynamics involving conical intersections.\cite{guo2024electronic} 
A similar compact form of HOPS, derived with bare ladder operators, has also been employed with MPS/MPO techniques to study many-body open quantum dynamics.\cite{flannigan2022many}

Thus far, applications of the cNMSSE have primarily focused on environments characterized by Debye-type spectral densities (SDs), for which the bath correlation function (BCF) can be analytically represented as a sum of exponential functions via Cauchy's residue theorem. 
However, for more general SDs, particularly those involving second- or higher-order poles, or those that are non-analytical and lack well-defined poles, the exponential basis may no longer be suitable, thereby necessitating alternative basis expansions.

This issue has been extensively explored within the framework of HEOM. 
In its original formulation, an exponential basis was employed to treat Debye-type SDs.\cite{tanimura1989time}
To extend its applicability to more general spectral forms, one common strategy is to approximate the SD using a sum of modes with first-order poles.
In this context, Meier and Tannor introduced a numerical fitting scheme based on a sum of Lorentzian modes.\cite{meier1999non}
Building upon this idea, Kleinekath\"ofer proposed a similar method incorporating both Debye and oscillatory Debye modes.\cite{kleinekathofer2004non}
This line of development was further advanced by Liu \emph{et al.}, who reformulated the HEOM framework to accommodate arbitrary SDs.\cite{liu2014reduced}
An alternative approach involves expanding the BCF directly using a general, non-exponential basis. 
A notable example is the generalized HEOM formulation introduced by Xu \emph{et al.},\cite{xu2005exact} which was later extended by Tang \emph{et al.} into the so-called extended HEOM framework, capable of treating a broader class of SDs.\cite{tang2015extended,duan2017zero,duan2017study,wang2019dynamical}
Additionaly, a decomposition strategy based on Chebyshev polynomials was proposed by Rahman and Kleinekath\"ofer. \cite{rahman2019chebyshev}
More recently, Ikeda and Scholes introduced a compact representation of extended HEOM, which reduces the dimensional hierarchy from $2K$ to $K$, where $K$ denotes the number of general basis functions.\cite{ikeda2020generalization}

While significant efforts have been devoted to extending the HEOM framework through non-exponential basis expansions, recent developments have also focused on improving the efficiency of exponential decompositions of BCFs.
For example, Xu \emph{et al.} proposed the barycentric spectral decomposition based on the AAA algorithm,\cite{nakatsukasa2018aaa,nakatsukasa2020algorithm} which provides accurate rational approximations for arbitrary SDs.\cite{xu2022taming,dan2023efficient}
In addtion, time-domain fitting strategies such as Prony's method,\cite{chen2022universal} as well as other recent schemes,\cite{takahashi2024high} have been employed to construct compact exponential representations of BCFs.

Within the context of NMSSE methods, we also note that non-exponential BCF expansions have been considered in the hierarchy of forward-backward stochastic Schr\"odinger equation (HFB-SSE) proposed by Ke and Zhao.\cite{ke2016hierarchy} 
However, the hierarchy structure in their approach must be carefully re-derived for each specific choice of basis function, which significantly complicates the treatment of general spectral forms.

In this work, we extend the cNMSSE formalism to accommodate arbitrary SDs using general basis functions, resulting in the extended cNMSSE. 
This extension retains the wavefunction-based formulation of the original cNMSSE while structurally resembling the extended HEOM framework proposed by Ikeda and Scholes.\cite{ikeda2020generalization}
Compared to the HFB-SSE, the extended cNMSSE offers a more concise and unified representation applicable to a wide range of spectral environments.
The accuracy and flexibility of the extended cNMSSE are demonstrated through four representative cases that employ non-exponential basis functions: a discrete SD, Ohmic SDs with exponential and algebraic cutoffs, and a critically damped Brownian SD.

The remainder of this paper is organized as follows. The derivation of the extended cNMSSE formalism with general basis functions is presented in section \ref{theory}. Numerical results for the aforementioned cases are reported in section \ref{numerical results}. Conclusions are given in section \ref{conclusion}.

\section{\label{theory} theoretical methods}

\subsection{\label{open quantum system} The open quantum system}
We consider a quantum system linearly coupled to a bath of harmonic oscillators, with the total Hamiltonian given by
\begin{align}
    \hat{H} &= \hat{H}_\mathrm{{S}} + \hat{H}_{\mathrm{B}} + \hat{H}_{\mathrm{SB}} \nonumber \\
            &= \hat{H}_\mathrm{{S}} + \sum_\lambda{(\frac{\hat{p}_\lambda^2}{2}+\frac{1}{2}\omega_\lambda^2\hat{q}_\lambda^2)} + f(q) \otimes \sum_\lambda{c_\lambda \hat{q}_\lambda}
\end{align}
where $\{\hat{q}_\lambda\}$ and $\{\hat{p}_\lambda\}$ denote the coordinates and momenta of the bath degrees of freedom, respectively, and $c_{\lambda}$ represents the coupling strength between the system and the $\lambda$th bath mode.

The SD characterizing the frequency-dependent system-bath coupling strength is defined as 
\begin{equation}\label{eq:SD}
    S (\omega)=\frac{\pi}{2} \sum_\lambda \frac{c_\lambda^2}{\omega_\lambda} \delta(\omega-\omega_\lambda) .
\end{equation}
In the time domain, the BCF characterizing the influence of the environment at temperature $T$ is expressed as 
\begin{align}\label{eq:BCF}
    \alpha(t)=& \int_0^\infty d\omega~ \frac{S(\omega)}{\pi} \left[\coth\left(\frac{\beta\omega}{2}\right) \cos\omega t - i \sin \omega t \right] \nonumber \\
    =&  \int_{-\infty}^{\infty}{ d\omega~  \frac{S(\omega)}{\pi} f_{\mathrm{Bose}}(\beta\omega)  e^{-i\omega t}}
\end{align}
with the inverse temperature $\beta=1/T$ and Bose distribution function $f_{\mathrm{Bose}}(x)=1/(1-e^{-x})$. 
The dynamics of system can be obtained by the reduced density matrix
\begin{equation}
    \rho(t) = \mathrm{Tr_B} \left[\rho_{\mathrm{tot}}(t)\right] ,
\end{equation}
where $\rho_{\mathrm{tot}}(t)$ is the total density matrix and $\mathrm{Tr_B}\left[\cdot\cdot\cdot\right]$ denotes the trace over all bath degrees of freedom. Throughout, we adopt natural units $\hbar=k_B=1$ for simplicity.

\subsection{\label{hops}The NMQSD and mixed
deterministic-stochastic framework}
Assuming that the system and the bath are initially uncorrelated, and that the bath is in thermal equilibrium, the initial total density matrix can be written as $\rho_{\mathrm{tot}}(0)=\rho_\mathrm{S}(0)\otimes\frac{e^{-\hat{H}_\mathrm{B}/T}}{Z_{\mathrm{B}}}$ with the partition function defined as $Z_{\mathrm{B}}=\mathrm{Tr_B}\left[e^{-\hat{H}_\mathrm{B}/T} \right]$.
The evolution of the system can be expressed using a double path integral formalism\cite{feynman1963theory}
\begin{equation}
    \rho_\mathrm{S}(t) = \int\mathcal{D}[q^+] \int\mathcal{D}[q^-] e^{i\{S_0[q^+]-S_0[q^-]\}} \mathcal{F}[q^+, q^-] \rho_\mathrm{S}(0) ,
\end{equation}
where $S_0[q^\pm]$ denotes the action associated with the system Hamiltonian $\hat{H}_\mathrm{S}$ on the forward ($q^+$) and backward ($q^-$) paths, respectively. 
The environmental influence is encoded in the Feynman-Vernon influence functional (IF), which takes the form: 
\begin{widetext}
\begin{equation}
    \mathcal{F}[q^+, q^-] = \exp\left\{-\int_0^t ds \int_0^s du 
    \begin{pmatrix} f[q^+(s)] & f[q^-(s)] \end{pmatrix}
    \begin{pmatrix} \alpha(s,u) & -\alpha^*(s,u) \\ -\alpha(s,u) & \alpha^*(s,u) \end{pmatrix}
    \begin{pmatrix} f[q^+(u)] \\ f[q^-(u)]   \end{pmatrix}
    \right\},
\end{equation}
where $\alpha(t,s)=\alpha(t-s)$. 
The forward and backward paths are coupled through the off-diagonal terms of the IF.
To introduce a stochastic representation, we define the adjusted BCF (aBCF) as $\tilde{\alpha}(t)=\alpha(t)-\alpha^1(t)$, and rewrite the IF in the following form:
\begin{equation}
    \mathcal{F}[q^+, q^-] = \exp(-\Phi_\mathrm{res}[q^+]-\Phi^*_\mathrm{res}[q^-])
    \exp\left\{-\int_0^t ds \int_0^s du 
    \begin{pmatrix} f[q^+(s)] & f[q^-(s)] \end{pmatrix}
    \begin{pmatrix} \alpha_1(s,u) & -\alpha^*(s,u) \\ -\alpha(s,u) & \alpha_1^*(s,u) \end{pmatrix}
    \begin{pmatrix} f[q^+(u)] \\ f[q^-(u)]   \end{pmatrix}
    \right\},
\end{equation}
where $\Phi_\mathrm{res}[q^\pm] =\int_0^t ds \int_0^s du f[q^\pm(s)]\tilde{\alpha}(s,u)f[q^\pm(u)]$ and $\alpha_1(t)=\alpha(t)-\tilde{\alpha}(t)$.
Applying the Hubbard-Stratonovich transformation,\cite{hubbard1959calculation} the IF can be decoupled into a stochastic average:
\begin{equation}
    \mathcal{F}[q^+, q^-] = \exp(-\Phi_\mathrm{res}[q^+]-\Phi^*_\mathrm{res}[q^-]) 
    \left\langle\exp(\int_0^tds~ Z_+(s)f[q^+(s)] - Z_-^*(s)f[q^-(s)]) \right\rangle ,
\end{equation}
where $Z_\pm(t)$ are complex stochastic processes satisfying the following statistical properties:
\begin{align}
    \left\langle Z_+(t) \right\rangle&=\left\langle Z_-(t) \right\rangle=0 , \\
    \left\langle Z_+(t)Z_+(s) \right\rangle&=\left\langle Z_-(t)Z_-(s) \right\rangle=\alpha_1(t,s) , \\
    \left\langle Z_+(t)Z_-^*(s) \right\rangle&= \alpha^*(t,s) .
\end{align}
Here, $\langle \cdots \rangle$ denotes averaging over the stochastic processes. 
Assuming the system is initially in a pure state $\rho_\mathrm{S}(0)=\ket{\psi_0}\bra{\psi_0}$, the stochastic wavefunction can be expressed as
\begin{equation}
    \ket{\psi_t^\pm(Z_\pm)} = \int\mathcal{D}[q^\pm] \exp\left\{ iS_0[q^\pm] - \Phi_\mathrm{res}[q^\pm] - i \int_0^t ds~ Z_\pm(s) f[q^\pm(s)] \right\}\ket{\psi_0}.
\end{equation}
Taking the time derivative yields the NMQSD equation:\cite{diosi1997non, diosi1998non}
\begin{equation}
\label{eq:NMQSD}
    \frac{d}{dt} \left|\psi_t^\pm\left(Z_\pm\right)\right\rangle = -i\hat{H}_{\mathrm{S}} \left|\psi_t^\pm\left(Z_\pm\right)\right\rangle - i Z_\pm(t) f(\hat{q}) \left|\psi_t^\pm\left(Z_\pm\right)\right\rangle - i f(\hat{q}) \int_0^t du~ \tilde{\alpha}(t,u)\frac{\delta \left|\psi_t^\pm\left(Z_\pm\right)\right\rangle}{\delta Z_\pm(u)}.
\end{equation}
\end{widetext}
Then the reduced density matrix $\rho_\mathrm{S}(t)$ can be reconstructed from the stochastic wavefunctions as
\begin{equation}
    \rho(t) = \mathbb{E} \left[\left|\psi_t^+\left(Z_+\right)\right\rangle \left\langle\psi_t^-\left(Z_-\right)\right| \right] ,
\end{equation}
where $\mathbb{E}\left[\cdot\cdot\cdot\right]$ denotes the averages over stochastic trajectories.

Depending on the choice of of the deterministic or stochastic part of the BCF, which yields different forms of the aBCF $\tilde{\alpha}(t)$, Eq.~\ref{eq:NMQSD} recovers various known forms of the NMQSD equation.
In the original formulation by Di{\'o}si and Strunz,\cite{diosi1997non,diosi1998non,strunz1999open} one sets $\alpha_1(t)=0$ and $Z_+(t)=Z_-(t)$, leading to $\tilde{\alpha}(t)=\alpha(t)$.
Alternatively, Song and Shi \emph{et al.} proposed a scheme in which $Z_+(t)=Z_-(t)$, and $\alpha_1(t) = \int_0^\infty d\omega \frac{S(\omega)}{\pi} \mathrm{csch}(\frac{\beta\omega}{2})\cos{\omega t} $, resulting in the following form of the aBCF\cite{song2016alternative}:
\begin{align}\label{eq:aBCF1}
    \tilde{\alpha}(t)=&\int_0^\infty d\omega~ \frac{S(\omega)}{\pi} \left[\tanh\left(\frac{\beta\omega}{4}\right) \cos\omega t - i \sin \omega t \right] \nonumber \\
    =& -\int_{-\infty}^{\infty}{ d\omega~  \frac{S(\omega)}{\pi} f_{\mathrm{Fermi}}\left(\frac{\beta\omega}{2}\right)  e^{i\omega t}} ,
\end{align}
where $f_\mathrm{Fermi}(x)=1/(1+e^x)$ is the Fermi distribution function.
Compared to the origin form in Eq.~\ref{eq:BCF}, this choice leads to a faster decay of the memory kernel, especially in the high-temperature regime.
However, both the Di{\'o}si-Strunz and the Song–Shi approaches involve Bose or Fermi distribution functions, whose Matsubara or Pad\'e spectral decomposition\cite{hu2010communication,hu2011pade} introduces a large number of modes at low temperatures.
Fano spectrum decomposition has been shown to be more efficient than these two methods in such cases. \cite{cui2019highly,zhang2020hierarchical}
An alternative strategy, proposed by Ke and Zhao,\cite{ke2016hierarchy,ke2017perturbation} circumvents this issue by setting $Z_+(t)\neq Z_-(t)$ and $\alpha_1(t)= \Reals[\alpha(t)]$, which results in a purely imaginary aBCF:
\begin{equation}\label{eq:aBCF2}
    \tilde{\alpha}(t)=-i\int_0^\infty d\omega~ \frac{S(\omega)}{\pi} \sin{\omega t}. 
\end{equation}
This formulation forms the basis of the HFB-SSE, where the exponential terms $\nu_k$ in Eq.~\ref{eq:exp_basis} can be directly obtained from the poles of the SD $S(\omega)$. 
To directly illustrate the failure of the original cNMSSE for certain spectral densities, particularly those involving higher-order poles, we adopt the last strategy in Section~\ref{numerical results}, which avoids the complications associated with Bose or Fermi distribution functions.
Nevertheless, we emphasize that the extended cNMSSE framework developed in this work is fully compatible with all three constructions of the aBCF discussed above.
A comprehensive discussion of these methods and their numerical performance can be found in Refs.~\onlinecite{wang2019hierarchical} and \onlinecite{wang2020hierarchical}. 

\subsection{\label{cNMSSE} The extended HOPS and cNMSSE method}
The functional derivative term in Eq.~\ref{eq:NMQSD} is efficiently evaluated within the framework of HOPS.\cite{suess2014hierarchy}
Unlike HEOM, which typically decomposes the BCF in Eq.\ref{eq:BCF} as a sum of exponentials, the NMSSE formulation offers greater flexibility in representing the aBCF\cite{song2016alternative,ke2016hierarchy}.
Departing from the exponential basis used in original HOPS, we employ a complete basis set $\{\phi_k(t)\}$ satisfying
\begin{equation} \label{eq:general_basis}
    \partial_t\phi_k(t)=\sum_{k'=1}^K \eta_{kk'} \phi_{k'}(t),
\end{equation}
where $\eta$ is a $K\times K$ coefficient matrix.
With this choice, the aBCF can be approximated as
\begin{equation}\label{eq:BCF_decomposition}
    \tilde{\alpha}(t) \approx \sum_{k=1}^{K} d_{k} \phi_k(t),
\end{equation}
where the complex coefficients ${d_k}$ depend on the specific SD and the selected basis functions.
We then define the auxiliary wavefunctions as:
\begin{align}
    \ket{\psi_\pm^\Bn} =& \int\mathcal{D}[q^\pm] \prod_{k=1}^K \left[- \int_0^tdu~ \phi_k(t-u) f[\hat{q}^\pm(u)]\right]^{n_k} \nonumber \\
    &\exp\left\{iS_0[q^\pm] - \Phi_\mathrm{res}[q^\pm] - i \int_0^t ds~ Z_\pm(s) f[q^\pm(s)] \right\} \ket{\psi_0},
\end{align}
where the superscript $\Bn=\left\{n_{1} \cdots n_{k} \cdots n_{K}\right\}$ consisting of non-negative integer indices. 
Differentiating $\ket{\psi_\pm^\Bn}$ with respect to time yields the extended HOPS equation:
\begin{align}
\label{eq:hops1}
    \frac{d}{dt} \ket{\psi_\pm^\Bn} = 
    &-i[\hat{H}_\mathrm{S}+Z_\pm(t)f(\hat{q})] \ket{\psi_\pm^\Bn}
    + f(\hat{q}) \sum_{k=1}^K d_k \ket{\psi_\pm^{\Bn+\Be_k}} \nonumber \\
    &- f(\hat{q}) \sum_{k=1}^K \phi_k(0) n_k \ket{\psi_\pm^{\Bn-\Be_k}} \nonumber \\
    &+ \sum_{k,k'=1}^K n_k \eta_{k,k'}  \ket{\psi_\pm^{\Bn-\Be_k+\Be_{k'}}},
\end{align}
where $\Be_k$ is a short-hand vector notation of $\left\{0, \cdots, 1_k, \cdots, 0\right\}$. 
For the special case of exponential basis functions,
\begin{equation} \label{eq:exp_basis}
    \phi_k(t)=e^{-\nu_k t},
\end{equation}
Eq.~\ref{eq:hops1} recovers the standard HOPS formulation:\cite{suess2014hierarchy}
\begin{align} \label{eq:hops3}
    \frac{d}{dt} \ket{\psi_\pm^\Bn} = &-i[\hat{H}_\mathrm{S}+Z_\pm(t)f(\hat{q})-i\sum_{k=1}^K\nu_kn_k] \ket{\psi_\pm^\Bn} \nonumber\\ &+ f(\hat{q}) \sum_{k=1}^K d_k \ket{\psi_\pm^{\Bn+\Be_k}} - f(\hat{q}) \sum_{k=1}^K n_k \ket{\psi_\pm^{\Bn-\Be_k}}.    
\end{align}

To enhance numerical stability at higher hierarchy levels, we normalize the auxiliary wavefunctions $\ket{\psi_\pm^\Bn} \rightarrow (\prod_{k=1}^Kn_k!)^{-1/2}\ket{\psi_\pm^\Bn}$, analogous to the rescaling approach in Ref.~\onlinecite{ikeda2020generalization}.
Under this transformation, Eq.~\ref{eq:hops1} is reformulated as
\begin{align}
\label{eq:hops2}
    \frac{d}{dt} \ket{\psi_\pm^\Bn} =
    &-i[\hat{H}_\mathrm{S}+Z_\pm(t)f(\hat{q})] \ket{\psi_\pm^\Bn}  \nonumber\\
    &+ f(\hat{q}) \sum_{k=1}^K d_k \sqrt{n_k+1} \ket{\psi_\pm^{\Bn+\Be_k}}  \nonumber\\
    &- f(\hat{q}) \sum_{k=1}^K \phi_k(0)\sqrt{n_k} \ket{\psi_\pm^{\Bn-\Be_k}}  \nonumber\\
    &+ \sum_{k,k'=1}^K \eta_{k,k'} \sqrt{n_k(n_{k'}+1))} \ket{\psi_\pm^{\Bn-\Be_k+\Be_{k'}}}.
\end{align}

Note that the original wavefunction in Eq.~\ref{eq:NMQSD} corresponds to the one with all the indices being zero, $\left|\psi_t^\pm\left(Z_\pm\right)\right\rangle=\psi_\pm^\mathbf{0}$. 
All higher-order auxiliary wavefunctions $\ket{\psi_\pm^\mathbf{n}}$ with $\mathbf{n}\ne \mathbf{0}$ are initialized to zero at $t=0$.
In practical simulations, the hierarchy is truncated at a finite order to form a closed set of equations. 
A common strategy is to impose a terminator condition of the form $\ket{\psi_\pm^{\Bn+\Be_k}}=0$, for sufficiently large $\mathbf{n}$.
Moreover, several truncation methods, such as triangular truncation,\cite{suess2014hierarchy} n-particle approximation, and n-mode approximation\cite{zhang2018flexible} can be applied to reduce the number of coupled equations.

To proceed, we define the pseudo-Fock states $\left|\Bn\right\rangle=\left|n_1\cdots n_k\cdots n_K\right\rangle$, with the orthonormal relation $\left\langle\Bn|\Bn'\right\rangle=\delta_{\Bn\Bn'}$. 
The creation ($\hat{b}_k^\dagger$) and annihilation ($\hat{b}_k$) operators act on these states according to
\begin{align}
   \hat{b}_k^\dagger \left|\Bn\right\rangle &= \sqrt{n_k+1} \left|\Bn+\Be_k\right\rangle ,\\   
   \hat{b}_k \left|\Bn\right\rangle &= \sqrt{n_k} \left|\Bn-\Be_k\right\rangle .
\end{align}
We now introduce an enlarged Hilbert space where the total state $\left|\Psi_t^\pm\right\rangle $ is represented as the sum of all tensor products of auxiliary wavefunctions and corresponding pseudo-Fock states, $ \left|\Psi_t^\pm\right\rangle = \sum_{\mathbf{n}} \ket{\psi_\pm^\Bn} \otimes \left|\Bn \right\rangle$. 
In this formulation, the extended HOPS equation in Eq.~\ref{eq:hops2} can be compactly rewritten as
\begin{equation}
\label{eq:cNMSSE}
    \partial_t \left|\Psi_t^\pm\right\rangle = -i \hat{H}_\mathrm{eff}^\pm\left|\Psi_t^\pm\right\rangle
\end{equation}
with the effective non-Hermitian stochastic Hamiltonian
\begin{align}
\label{eq:effective stochastic Hamiltonian1}
    \hat{H}_{\mathrm{eff}}^\pm &= \hat{H}_{\mathrm{S}}  + Z_\pm(t) f(\hat{q}) - if(\hat{q}) \sum_{k=1}^K \phi_k(0) \hat{b}_k^\dagger \nonumber\\
    &+ if(\hat{q})\sum_{k=1}^K d_k\hat{b}_k + i \sum_{k,k'=1}^K \eta_{kk'}\hat{b}_k^\dagger \hat{b}_{k'}.
\end{align}
In the original cNMSSE formulation,\cite{gao2022non} the basis functions are exponentials such that $\eta_{kk'}=-\nu_k \delta_{kk'}$, and Eq.~\ref{eq:effective stochastic Hamiltonian1} reduces to
\begin{align}
\label{eq:effective stochastic Hamiltonian2}
    \hat{H}_{\mathrm{eff}}^\pm &= \hat{H}_{\mathrm{S}}  + Z_\pm(t) f(\hat{q}) - i \sum_{k=1}^K \nu_k\hat{b}_k^\dagger \hat{b}_k \nonumber\\
    &+ if(\hat{q})  \sum_{k=1}^K (d_k\hat{b}_k - \hat{b}_k^\dagger) .
\end{align}
Adopting a rescaling of the auxiliary wavefunctions $\ket{\psi_\pm^\Bn} \rightarrow (\prod_{k=1}^Kn_k!d_k^{n_k})^{-1/2}\ket{\psi_\pm^\Bn}$\cite{shi2009efficient} can be further simplify the form of the interaction:\cite{guo2024electronic,gao2022non}
\begin{align}
\label{eq:effective stochastic Hamiltonian3}
    \hat{H}_{\mathrm{eff}}^\pm &= \hat{H}_{\mathrm{S}}  + Z_\pm(t) f(\hat{q}) - i \sum_{k=1}^K \nu_k\hat{b}_k^\dagger \hat{b}_k \nonumber\\
    &+ if(\hat{q})  \sum_{k=1}^K \sqrt{d_k} (\hat{b}_k - \hat{b}_k^\dagger) ,
\end{align}
which describes a system linearly coupled to a set of non-interacting quanum harmonic oscillators characterized by complex frequencies $\{\nu_k\}$. 
In contrast, for general basis functions, the inter-mode coupling arises through the off-diagonal elements $\eta_{kk'} (k\ne k')$,  indicating interactions between the pseudo-modes in the enlarged space.
The system wavefunction at time $t$ is recovered from the ground state of $\left|\Psi_t^\pm\right\rangle$ via $\ket{\psi_\pm^\Bn}=\left\langle \mathbf{0} |\Psi_t^\pm \right\rangle$.
In practical implementations, the pseudo-Fock space must be truncated at a suitable maximum occupation number $n_\mathrm{max}$ in each mode to ensure numerical convergence. This truncation corresponds to limiting the hierarchical depth in the HOPS formalism.

\section{\label{numerical results} Numerical Results}
We now demonstrate the extended cNMSSE (Eq.~\ref{eq:effective stochastic Hamiltonian1}) using four representative examples that employ non-exponential basis functions.
To enable direct comparison with the original cNMSSE (Eq.~\ref{eq:effective stochastic Hamiltonian3}) results, we adopt the aBCF defined in Eq.~\ref{eq:aBCF2}, where the poles arise only from the SD $S(\omega)$.
Numerical calculations were performed within the MPS framework using the renormalizer package.\cite{ren2018time,ren2020mpo,li2020numerical} 

\subsection{Discrete SD}
In this section, we investigate spin-boson dynamics under a discrete SD.\cite{ke2016hierarchy}
The system Hamiltonian for the spin-boson model (SBM) is given by $\hat{H}_\mathrm{S}=\epsilon\hat{\sigma}_z+\Delta\hat{\sigma}_x$, where $\hat{\sigma}_z=\ket{1}\bra{1}-\ket{2}\bra{2}$ and $\hat{\sigma}_x=\ket{1}\bra{2}+\ket{2}\bra{1}$. 
The system-bath coupling operator is chosen as $f(\hat{q})=\hat{\sigma}_z$, and the spin is initially prepared in the state $\ket{1}$.

For the discrete SD defined in Eq.~\ref{eq:SD}, the corresponding aBCF in Eq.~\ref{eq:aBCF2} can be expressed analytically as:
\begin{equation}
    \tilde{\alpha}(t) = -i\sum_k \frac{c_k^2}{2\omega_k} \sin{\omega_k t} 
    = -\sum_k \frac{c_k^2}{4\omega_k} \left(e^{i\omega_kt}- e^{-i\omega_kt} \right).
\end{equation}
Accordingly, the complete non-exponential basis set can be chosen as $\phi_k(t) = \{\sin{\omega_kt}, \cos{\omega_k t}\}$, which satisfies the differential relation
\begin{equation}
    \eta = \eta_1 \oplus \cdots \oplus \eta_k \oplus \cdots \oplus \eta_K, ~
    \eta_k = \begin{pmatrix}
        0 & \omega_k \\ -\omega_k & 0
    \end{pmatrix}.
\end{equation}

\begin{figure}
    \centering
    \includegraphics[width=8.6 cm]{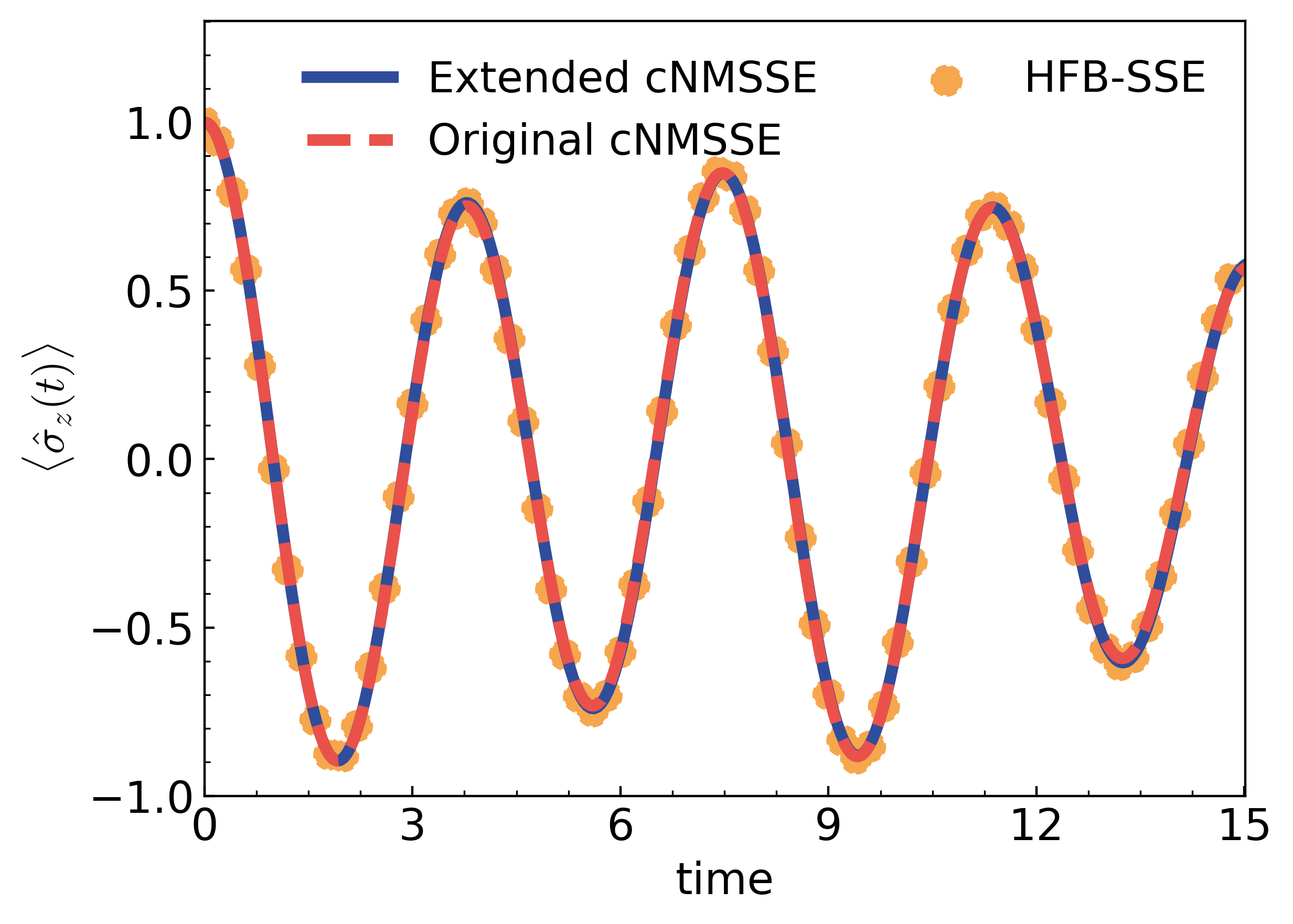}
    \caption{
    Population dynamics of the SBM with the discrete SD defined in Eq.~\ref{eq:SD}. 
    Simulation parameters are: $\epsilon=0$, $\Delta=0.5$, $c_k=0.2$, $\omega_k=1.0$ and $\beta=1.0$. 
    Results of both the extended and original cNMSSE methods are averaged over $10^4$ trajectories.}
    \label{fig:discrete_SD}
\end{figure}

Fig.~\ref{fig:discrete_SD} presents the population dynamics obtained using both the extended and original cNMSSE methods.
Simulation parameters are chosen as $\epsilon=0$, $\Delta=0.5$, $c_k=0.2$, $\omega_k=1.0$ and $\beta=1.0$, consistent with those used in Ref.~\onlinecite{ke2016hierarchy} for direct comparison.
In both simulations, two effective modes are included, each truncated at $n_\mathrm{max}=3$.
The results are averaged over $10^4$ stochastic trajectories.
For validation, we compare our simulations with results obtained from the HFB-SSE method.\cite{ke2016hierarchy}

\subsection{Ohmic SD with an exponential cutoff}
In this section, we investigate spin-boson dynamics for the Ohmic SD with an exponential cutoff.\cite{ke2016hierarchy} 
For this case, we adopt the decomposition strategy proposed by Meier and Tannor,\cite{meier1999non} which approximates the spectral density as a sum of Lorentzian modes:
\begin{equation}
    J(\omega) = \sum_k p_k \frac{\omega}{[(\omega+\Omega_k)^2+\Gamma_k^2][(\omega-\Omega_k)^2+\Gamma_k^2]},
\end{equation}
where the parameters $\{p_k,\Omega_k,\Gamma_k\}$ are determined via numerical fitting.
With this decomposition, the aBCF defined in Eq.~\ref{eq:aBCF2} can be expressed analytically as 
\begin{align}
    \tilde{\alpha}(t) =& -i\sum_k \frac{p_k e^{-\Gamma_k t}}{4\Omega_k\Gamma_k}\sin{\Omega_k t}  \nonumber\\
    =& \sum_k \frac{p_k}{8\Omega_k\Gamma_k} \left[e^{-(\Gamma_k+i\Omega_k)t} - e^{-(\Gamma_k-i\Omega_k)t} \right]
\end{align}
Accordingly, we employ the non-exponential basis functions $\phi_k(t) = \{e^{-\Gamma_k t}\sin{\Omega_kt}, e^{-\Gamma_k t}\cos{\Omega_k t}\}$, which satisfy the differential relation:
\begin{equation}
    \eta = \eta_1 \oplus \cdots \oplus \eta_k \oplus \cdots \oplus \eta_K, ~
    \eta_k = \begin{pmatrix}
        -\Gamma_k & \Omega_k \\ -\Omega_k & -\Gamma_k
    \end{pmatrix}.
\end{equation}

Specifically, we consider the form: 
\begin{equation}\label{eq:Ohmic_SD_exp1}
    S(\omega) = \alpha\omega e^{-\omega/\omega_c},
\end{equation}
where $\alpha$ represents the system-bath coupling strength and $\omega_c$ is the cutoff frequency that determines the exponential decay of the SD at high frequencies.
Three Lorentzian modes are used in the Meier–Tannor decomposition, and the fitting parameters are summarized in Table.~\ref{tab:label1}. 

\begin{table}[]
    \caption{
    Fitting parameters for the Meier-Tannor decomposition of the Ohmic SD with an exponential cutoff, as defined in Eq.~\ref{eq:Ohmic_SD_exp1}.}
    \begin{ruledtabular}
    \begin{tabular}{cccc}
         Mode & $p_k/(\alpha\omega_c^4)$ & $\Omega_k/\omega_c$ & $\Gamma_k/\omega_c$ \\ \hline
         1 & 12.0677 & 0.2378 & 2.2593 \\
         2 & -19.9762 & 0.0888 & 5.4377 \\
         3 & 0.1834 & 0.0482 & 0.8099 \\
    \end{tabular}
    \label{tab:label1}
    \end{ruledtabular}
\end{table}

\begin{figure}
    \centering
    \includegraphics[width=8.6 cm]{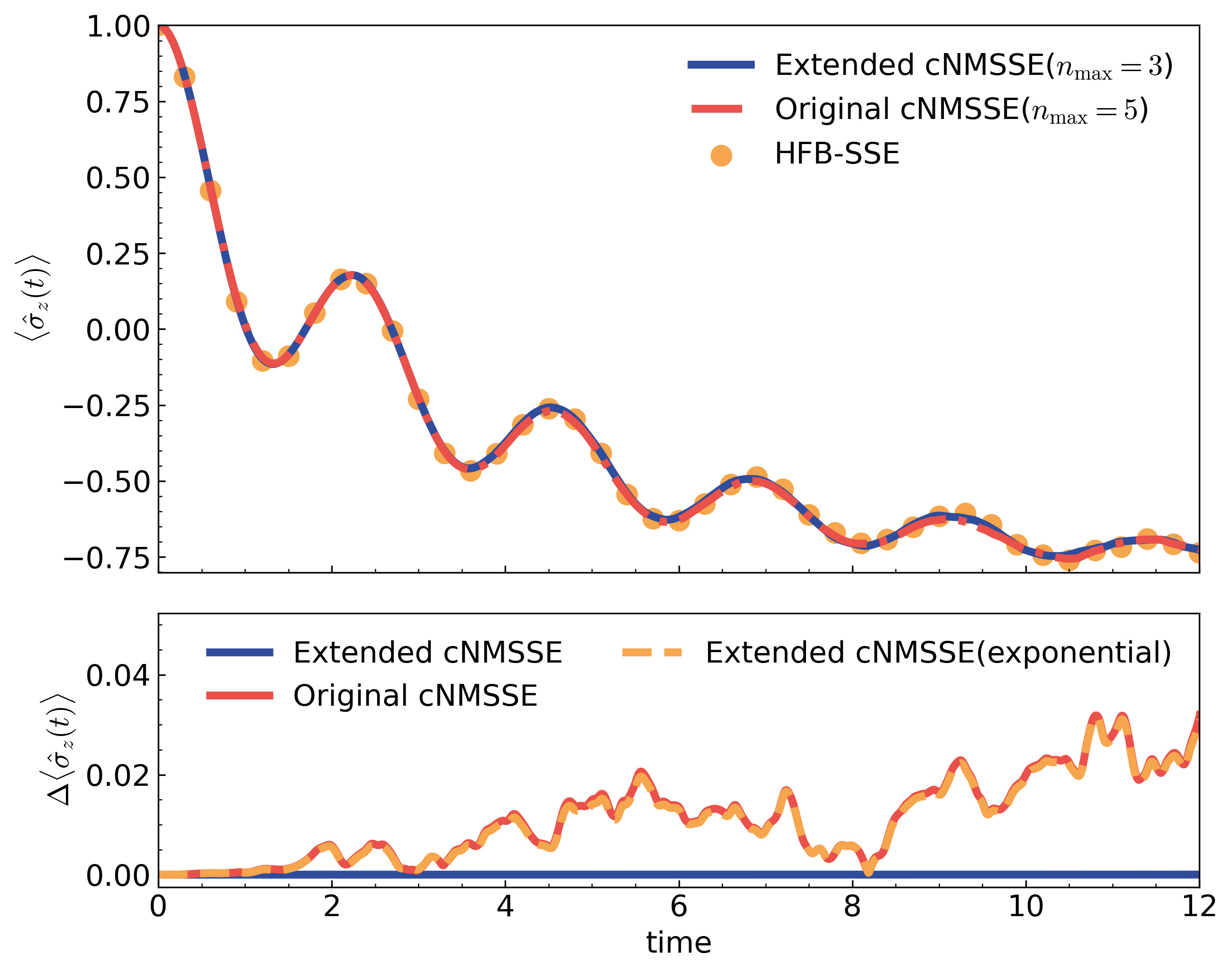}
    \caption{
    Top panel:Population dynamics of the SBM with the Ohmic SD with an exponential cutoff defined in Eq.~\ref{eq:Ohmic_SD_exp1}. 
    Simulation parameters: $\epsilon=\Delta=1.0$, $\alpha=0.157$, $\omega_c=7.5$ and $\beta=5.0$.
    Results of the extended and original cNMSSE methods are averaged over $2\times10^5$ trajectories.
    Bottom panel: A representative stochastic trajectory computed using three methods with $n_\mathrm{max}=3$: 
    (i) extended cNMSSE with non-exponential basis functions, 
    (ii) original cNMSSE, and (iii) extended cNMSSE with exponential basis functions. 
    }
    \label{fig:Ohmic_SD_exp}
\end{figure}

We perform simulations with parameters $\epsilon=\Delta=1.0$, $\alpha=0.157$, $\omega_c=7.5$ and $\beta=5.0$, following the benchmark setup in Ref.~\onlinecite{ke2016hierarchy}. 
The results are shown in Fig.~\ref{fig:Ohmic_SD_exp}.
Both the extended and original cNMSSE methods employ $6$ effective modes, with $n_\mathrm{max}=3$ for the extended cNMSSE and $n_\mathrm{max}=5$ for the original one to ensure numerical convergence.

The top panel shows the population dynamics over $2\times10^5$ trajectories, demonstrating excellent agreement between both methods and with the HFB-SSE reference. \cite{ke2016hierarchy}
The bottom panel displays a representative stochastic trajectory using three different formulations with $n_\mathrm{max}=3$: (i) the extended cNMSSE with non-exponential basis functions Eq.~\ref{eq:effective stochastic Hamiltonian1}, (ii) the original cNMSSE Eq.~\ref{eq:effective stochastic Hamiltonian3}, and (iii) the extended cNMSSE employing exponential basis functions Eq.~\ref{eq:effective stochastic Hamiltonian2}.
The latter formulation is employed to eliminate the influence of varying rescaling schemes.
Notably, even with a lower phonon cutoff, the extended formulation captures the correct dynamics. 
This is attributed to inter-mode couplings that redistribute bath excitations, thereby reducing the required occupation number in individual modes.

\subsection{Ohmic SD with an algebraic cutoff}
We next investigate spin-boson dynamics under the Ohmic SD with an algebraic cutoff, using the example introduced in Ref.~\onlinecite{ke2016hierarchy}:
\begin{equation}
\label{eq:Ohmic_SD_alg}
    S(\omega)=\frac{2\pi\alpha\omega}{(1+(\omega/\omega_c)^2)^2},
\end{equation}
where $\alpha$ is the dimensionless Kondo parameter that characterizes the system-bath coupling strength, and $\omega_c$ is the cutoff frequency controlling the effective bandwidth of bath.
The corresponding aBCF in Eq.~\ref{eq:aBCF2} can be analytically expressed as:
\begin{equation}
    \tilde{\alpha}(t) = -i\frac{\pi\alpha\omega_c^3}{2}te^{-\omega_c t}.
\end{equation}
This form naturally suggests a non-exponential basis set $\phi_k(t) = \{t e^{-\omega_ct}, e^{-\omega_ct}\}$, which satisfies
\begin{equation}
    \eta = \begin{pmatrix}
        -\omega_c & 1 \\ 0 & -\omega_c
    \end{pmatrix}.
\end{equation}
Since the algebraic form of the spectral density in Eq.~\ref{eq:Ohmic_SD_alg} leads to second-order poles and does not admit an analytical exponential basis, we employ the barycentric spectral decomposition scheme\cite{xu2022taming,dan2023efficient} to approximate $\tilde{\alpha}(t)$ for use in the original cNMSSE.

\begin{figure}
    \centering
    \includegraphics[width=8.6 cm]{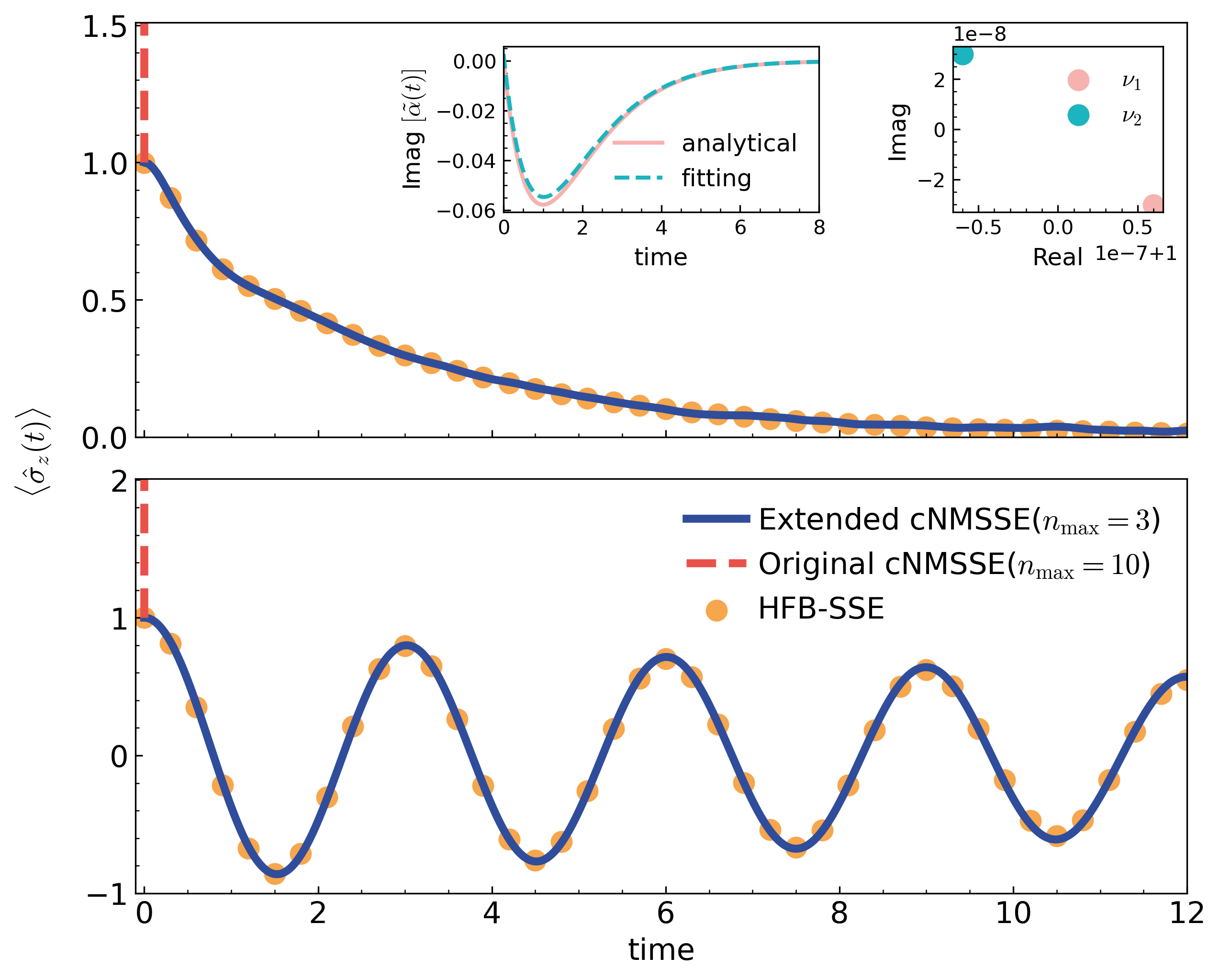}
    \caption{
    Population dynamics of the SBM with the Ohmic spectral density with an algebraic cutoff defined in Eq.~\ref{eq:Ohmic_SD_alg}. 
    Simulation parameters: $\epsilon=0$, $\Delta=1.0$, $\alpha=0.1$, and $\omega_c=1$.
    Top panel: high-temperature regime ($\beta=0.02$). 
    Bottom panel: low-temperature regime ($\beta=50$). 
    Both extended and original cNMSSE results are averaged over $10^5$ trajectories.
    Insets (top panel): (left) comparison of exact and fitted aBCFs; 
    (right) two exponential components $\nu_k$ from the barycentric spectral decomposition.
    }
    \label{fig:Ohmic_SD_alge1}
\end{figure}

Fig.~\ref{fig:Ohmic_SD_alge1} shows population dynamics obtained using both extended and original cNMSSE methods for two temperature regimes: high temperature ($\beta=0.02$) and low temperature ($\beta=50$).
The simulations are performed with $\epsilon=0$, $\Delta=1.0$, $\alpha=0.1$ and $\omega_c=1$, using two effective modes with $n_\mathrm{max}=3$, and $n_\mathrm{max}=10$, respectively.
Results are averaged over $10^5$ stochastic trajectories.

The extended cNMSSE accurately captures the system dynamics and is consistent with HFB-SSE results.\cite{ke2016hierarchy}
In contrast, the original cNMSSE fails in both temperature regimes.
This failure arises from numerical instabilities caused by the near-degenerate poles introduced by the algebraic structure of $S(\omega)$ in the barycentric decomposition.

\begin{figure}
    \centering
    \includegraphics[width=8.6 cm]{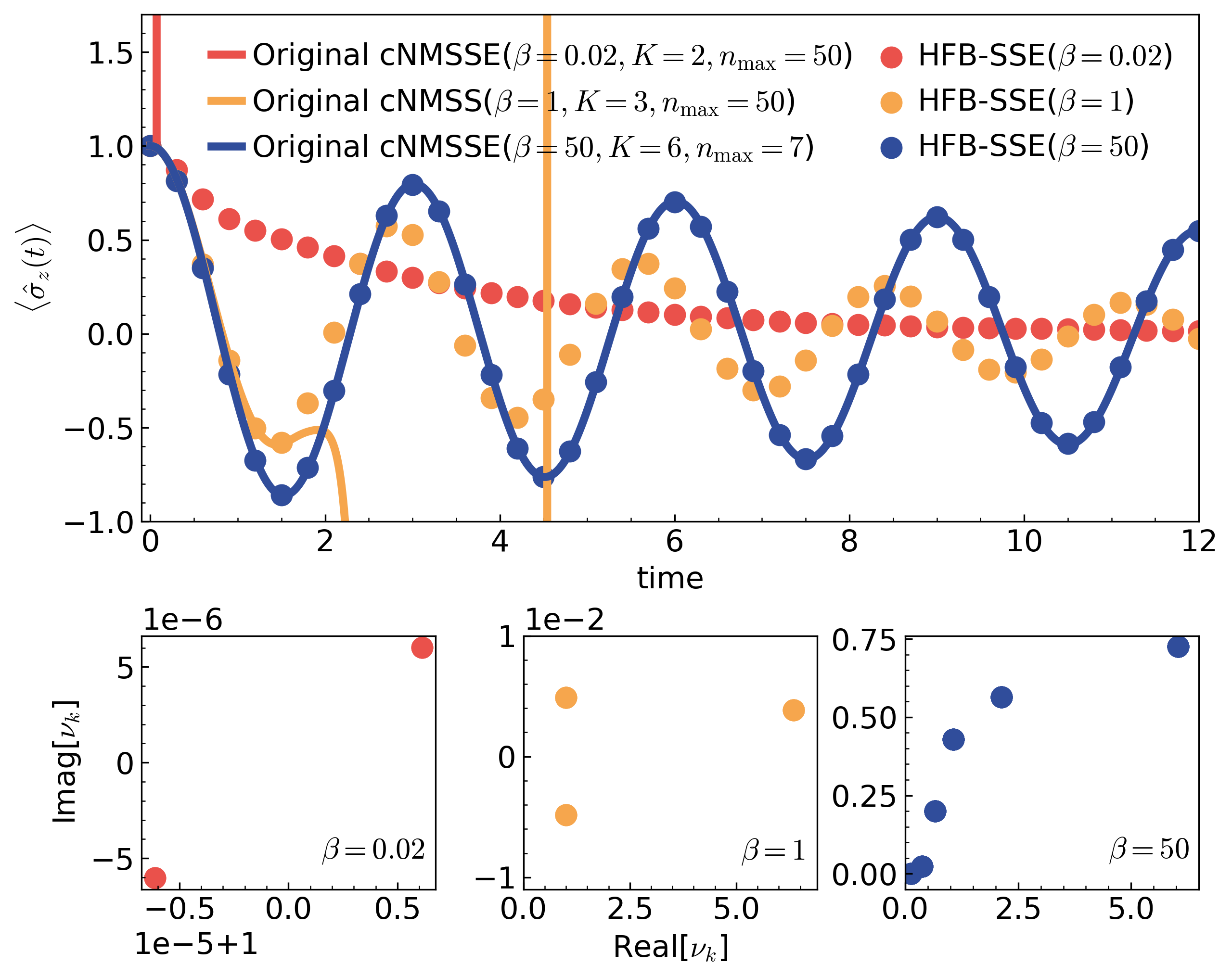}
    \caption{
    Top panel: Population dynamics of the SBM with the Ohmic SD with an algebraic cutoff as defined in Eq.~\ref{eq:Ohmic_SD_alg}, using the original cNMSSE with the temperature-dependent aBCF from Eq.\ref{eq:aBCF1}. 
    Simulation parameters: $\epsilon=0$, $\Delta=1.0$, $\alpha=0.1$, and $\omega_c=1$. 
    Results are averaged over $10^5$ trajectories. 
    Bottom panel: Exponential components $\nu_k$ obtained via barycentric decomposition at different temperatures: (left) $\beta=0.02$, (center) $\beta=1$, and (right) $\beta=50$.
    }
    \label{fig:Ohmic_SD_alge2}
\end{figure}

To further explore the role of temperature in barycentric spectral decomposition, we consider the aBCF form in Eq.\ref{eq:aBCF1}, which depends explicitly on temperature via the Fermi distribution function $f_\mathrm{Fermi}$.
Figure\ref{fig:Ohmic_SD_alge2} shows that, at high temperature ($\beta=0.02$), the decomposition produces degenerate poles that again destabilize the original cNMSSE simulation.
As the temperature decreases, the Fermi distribution function term dominates, and the decomposition yields more diverse and less-degenerate exponential components, thereby improving numerical stability.
This illustrates that, in environments where the aBCF is predominantly determined by the SD $S(\omega)$, non-exponential basis functions may provide a more natural and stable representation than exponential expansions, especially when higher-order poles are involved.

\subsection{Critically damped Brownian SD}
We now investigate exciton/electron transfer models with a Brownian spectral density.\cite{ikeda2020generalization} 
We consider a simplified model described by the Hamiltonian
\begin{equation}
    \hat{H}_\mathrm{S} = E_\mathrm{D} \ket{D}\bra{D} + (E_A+\lambda) \ket{A}\bra{A} + J(\ket{D}\bra{A}+\ket{A}\bra{D}) ,
\end{equation}
where $\ket{D}$ and $\ket{A}$ represent the reactant state and product state of the transfer process, respectively.
And $\lambda$ denotes the reorganization energy associated with the $\ket{D} \rightarrow \ket{A}$ transition. 
The system is initialized in the excited donor state $\ket{D}$.
The environment is described by the Brownian SD
\begin{equation}
\label{eq:Brownian_SD}
    S(\omega) = 2\lambda \frac{\zeta\omega_0^2\omega}{(\omega^2-\omega_0^2)^2+\zeta^2\omega^2}
\end{equation}
where $\omega_0$ is the characteristic bath mode frequency and $\zeta$ is the damping constant. 
The dynamical regime of the bath is classified by the relation between $\zeta$ and $\omega_0$: underdamped for $\zeta<2\omega_0$, critically damped for $\zeta=2\omega_0$, and overdamped for $\zeta>2\omega_0$.

In the underdamped and overdamped regimes, the aBCF in Eq.~\ref{eq:aBCF2} admits an exponential decomposition:
\begin{equation}\label{eq:Brownian_exp_basis}
    \tilde{\alpha}(t) = \frac{\lambda\omega_0^2}{2\omega_1} e^{-\gamma_+ t} 
    - \frac{\lambda\omega_0^2}{2\omega_1} e^{-\gamma_- t} ,
\end{equation}
where $\gamma_\pm=\zeta/2\pm i\omega_1$, and $\omega_1=\sqrt{\omega_0^2-\zeta^2/4}$.
This form enables the direct application of the original cNMSSE.

\begin{figure}
    \centering
    \includegraphics[width=8.6 cm]{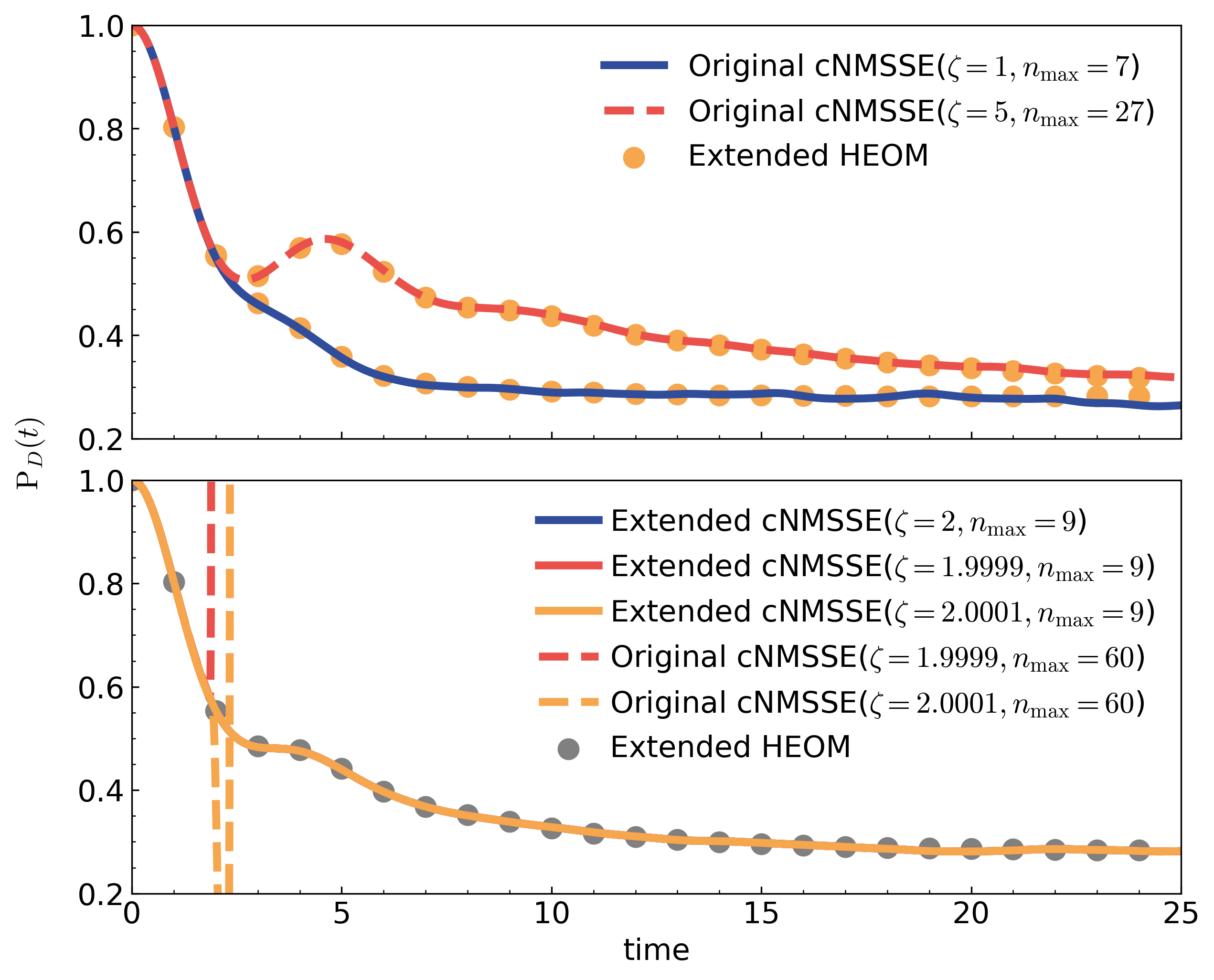}
    \caption{
    Donor population dynamics for the electron/exciton transfer model with the Brownian SD defined in Eq.~\ref{eq:Brownian_SD}. 
    Simulation parameters: $E_D-E_A=1.0$, $J=0.5$, $\omega_0=1.0$, $\lambda=1.0$ and $\beta=1.0$. 
    Top panel: Underdamped ($\zeta=1$) and overdamped ($\zeta=5$) regimes.
    Bottom panel: (Near) critically damped case.
    Results of the extended and original cNMSSE methods are averaged over $2\times10^4$ trajectories.
    }
    \label{fig:Brownian_SD1}
\end{figure}
However, in the critically damped case, the two first-order poles of the SD coalesce into a second-order pole.
This degeneracy causes the two exponential terms in Eq.~\ref{eq:Brownian_exp_basis} to become linearly dependent, leading to numerical instabilities in methods that rely on exponential basis expansions, including the original HEOM and cNMSSE.\cite{ikeda2020generalization}
To resolve this issue, we employ a non-exponential basis set that remains linearly independent at critical damping:\cite{ikeda2020generalization}
\begin{align}
    \phi_p(t) =& -\frac{\omega_0}{\omega_1} \sin{(\omega_1 t)} e^{-\zeta t/2}, \nonumber \\
    \phi_q(t) =& \left(\frac{\zeta}{2\omega_1} \sin{\omega_1t} + \cos{\omega_1t} \right) e^{-\zeta t/2},
\end{align}
which satisfies
\begin{equation}
    \eta = \begin{pmatrix}
        -\zeta & -\omega_0  \\
        \omega_0 & 0  
    \end{pmatrix}.
\end{equation}
Using this basis, the aBCF becomes
\begin{equation}
    \tilde{\alpha}(t) = i\lambda\omega_0 \phi_p(t),
\end{equation}
which ensures numerical stability and accuracy in the extended cNMSSE simulations.

\FloatBarrier
Fig.~\ref{fig:Brownian_SD1} presents the donor population dynamics for $E_D-E_A=1.0$, $J=0.5$, $\lambda=1.0$, $\omega_0=1.0$ and $\beta=1.0$, covering underdamped ($\zeta=1$), critically damped ($\zeta=2$), and overdamped ($\zeta=5$) regimes.
In each cases, two effective modes are employed, with $n_\mathrm{max}=7$, $9$ and $27$, respectively.
Results are averaged over $2\times10^4$ stochastic trajectories and benchmarked against the extended HEOM. \cite{ikeda2020generalization}

In the underdamped and overdamped regimes, both the original and extended cNMSSE employ the same exponential basis and yield identical, stable results. 
However, near the critical damping condition, the degeneracy of exponential basis functions renders the original cNMSSE numerically unstable. 
In contrast, the extended cNMSSE, which incorporates non-exponential basis sets, remains stable and accurate. 
This highlights the necessity of using general basis functions when the SD exhibit second-order (or higher-order) poles.

\section{\label{conclusion} Conclusion}
In this work, we have developed an extended cNMSSE method  employing general basis functions, advancing beyond the original cNMSSE constrained by exponential decompositions of aBCF.
This extension significantly broadens the applicability of the original cNMSSE by enabling accurate and efficient simulations of open quantum system dynamics under arbitrary SDs, including those beyond the Debye-type forms.
Such flexibility is particularly crucial for SDs with second- or higher-order poles, where exponential expansions either become inaccurate or inapplicable.
In such cases, the extended cNMSSE avoids the numerical instabilities caused by degenerate exponential terms and offers a more general, robust, and scalable framework for capturing non-Markovian memory effects.

The accuracy and efficiency of the extended cNMSSE are demonstrated through benchmark simulations on four representative models: a discrete SD, Ohmic SDs with exponential and algebraic cutoffs, and a critically damped Brownian SD. 
In all cases, excellent agreement is observed with results from the HFB-SSE and extended HEOM methods.
Notably, the aBCFs in all benchmark cases are analytically expanded using non-exponential basis functions. 
For more complex or structured SDs, identifying appropriate basis sets or developing efficient numerical decomposition strategies remains an open direction for future research.

\begin{acknowledgments}
The authors thank Y.Ke for sharing her code on HFB-SSE, and T.Ikeda for valuable discussions regarding the PyHEOM package. 
X.G.~acknowledge the support from the National Natural Science Foundation of China (Grant Nos. 22273122 and T2350009), the Guangdong Provincial Natural Science Foundation (Grant No.2024A1515011504), and computational resources and services provided by the national supercomputer center in Guangzhou. 
\end{acknowledgments}

\FloatBarrier

\nocite{*}
\bibliography{main}
\end{document}